\documentclass[aip,cha,amsmath,amssymb,reprint]{revtex4-1}

\usepackage{graphicx}
\usepackage{dcolumn}
\usepackage{bm}

\usepackage[utf8]{inputenc}
\usepackage[T1]{fontenc}
\usepackage{mathptmx}
\usepackage{etoolbox}
\usepackage{amssymb}
\usepackage{amsmath}
\usepackage{amsthm}
\usepackage{colortbl}
\usepackage[table,xcdraw]{xcolor}
\usepackage{amsmath}
\renewcommand{\arraystretch}{1.5}

\makeatletter
\def\@email#1#2{%
 \endgroup
 \patchcmd{\titleblock@produce}
  {\frontmatter@RRAPformat}
  {\frontmatter@RRAPformat{\produce@RRAP{*#1\href{mailto:#2}{#2}}}\frontmatter@RRAPformat}
  {}{}
}%
\makeatother
\begin{document}

\preprint{AIP/123-QED}

\title[Stability of the 1d swarmalator model in the continuum limit]{Stability of the 1d swarmalator model in the continuum limit}

\author{Kevin O'Keeffe}
 \altaffiliation[Starling Research Institute]{Starling Research Institute}

\date{\today}

\begin{abstract}
We study the 1d swarmalator model in the continuum limit. We examine the stability of its collective states which have compact support: synchrony, where the swarmalators lie in two sync dots (zero dimensional support), and the phase wave, where the swarmalators line up in a ring with uniformly spaced positions and phases (one dimensional support). The compact support imposes analytic difficulties that occur in many other swarmalator models and thus is blocking progress in the field. We show how to overcome this difficulty, deriving the two states' stability spectra exactly.
\end{abstract}

\maketitle

\begin{quotation}
Synchronization is a classic topic in nonlinear science. Ordinarily, the focus is on the oscillators' phase dynamics; their spatial degrees of freedom are either ignored entirely, as in mean field models, or are dynamically inactive, as in lattice models where the oscillators are pinned at fixed locations in space. By contrast, a new wave of work considers oscillators which are mobile, swarming through space as they synchronize in time. One barrier in analyzing so-called swarmalators is that their steady state patterns are often rings, annuli or disks; linearizing around states like these which have compact support leads to eigenvalue equations with generalized functions (delta functions and their relatives) whose analysis is non-standard. We present a case study of the 1d swarmalator model where such eigenvalue equations can be solved exactly. Specifically, we compute the stability spectrum of the model's sync and phase wave states in the continuum limit. These states are 1d analogues of several real world swarmalators states, such as vortex arrays of sperm \cite{riedel2005self} and sync crystals of Janus particles \cite{yan2015rotating}, and thus are representative of generic swarmalator phenomena. Our results help overcome a common obstacle in the study of swarmalators, and in that sense advance the field. They may also be useful in other parts of nonlinear science where compactly supported patterns arise, such as bacterial aggregation \cite{tsimring1995aggregation}, developmental embryology \cite{tsiairis2016self},  or colloidal physics \cite{trivedi2020room}. 
\end{quotation}

\section{Introduction}
Swarmalators are mobile oscillators whose internal phase dynamics and external motions are coupled \cite{o2017oscillators}. They model the many real world systems which self-assemble in space and self-synchronize in time interdependently, such as vinegar eels \cite{quillen2021metachronal}, tree frogs \cite{aihara2014spatio}, magnetic domain walls \cite{hrabec2018velocity}, colloidal micromotors \cite{yan2012linking,liu2021activity,zhang2020reconfigurable}, embryonic cells \cite{tsiairis2016self}, and robotic swarms \cite{barcis2019robots,barcis2020sandsbots} 

This paper is about the 1d swarmalator model \cite{o2022collective,yoon2022sync} where the units move on a 1d periodic domain
\begin{align}
    \dot{x}_i &= \nu_i' + \frac{J'}{N} \sum_j \sin(x_j - x_i) \cos(\theta_j - \theta_i) \\
    \dot{\theta}_i &= \omega_i' + \frac{K'}{N} \sum_j \sin(\theta_j - \theta_i) \cos(x_j - x_i) 
\end{align}
Here $x_i, \theta_i \in S^1$ are the position and phase of the $i$-th swarmalator,  ($\nu_i', \omega_i$') its natural frequencies, and $(J',K')$ the coupling constants. The phase dynamics model synchronization (the Kuramoto sine term) which depends on distance (the new cosine term). Conversely, the spatial dynamics model swarming/aggregation which depends on phase similarity. 

The simple, symmetric form of the 1d swarmalator model makes it one of the few models of mobile oscillators that is tractable. It has been used to shed analytic light on swarmalators with disordered coupling \cite{o2022swarmalators,hao2023attractive}, phase frustration \cite{lizarraga2023synchronization}, random pinning \cite{sar2023pinning,sar2024solvable,sar2023swarmalators}, external forcing \cite{anwar2024forced}, three-body interactions \cite{anwar2024collective}, and thermal noise \cite{hong2023swarmalators}.

\begin{figure}[t!]
\centering
\includegraphics[width = \columnwidth]{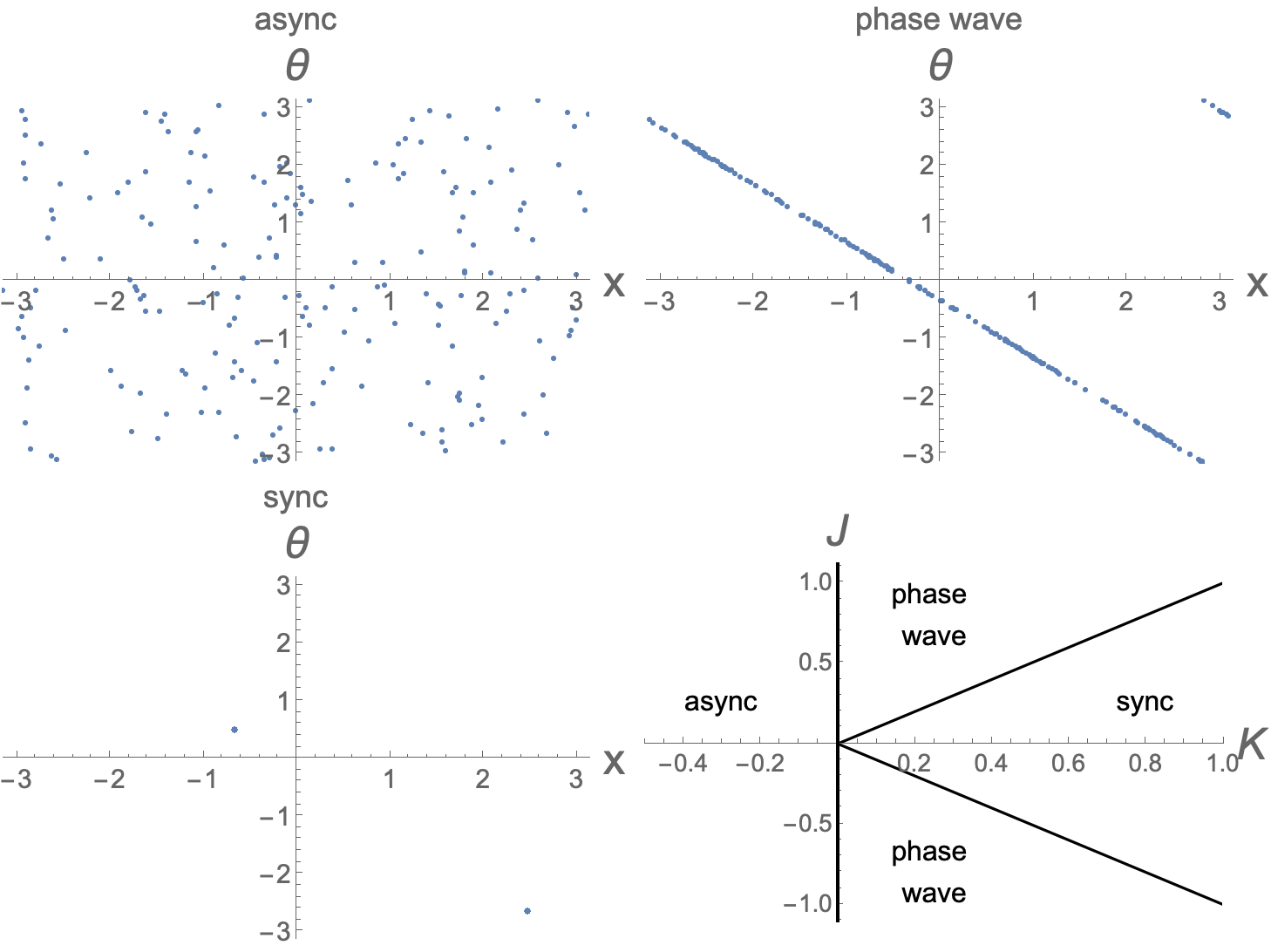}
\caption{(a)-(c) Collective states of 1d swarmalator model in the $N \rightarrow \infty$ limit for $(J',K') = (1,-2), (1,-0.5, 1,2))$. Here $(dt,T,N) = (0.1,100,200)$ and initial condition were drawn uniformly at random from $[-\pi,\pi]^2$. (d) Stability diagram in the $(J,K)$ plane.}
\label{states-identical}
\end{figure}

Yet even within the simplified case of identical swarmalators unanswered questions remain. For example, the stability of its two compactly supported states in the continuum limit are unknown. Figure~\ref{states-identical}(a) plots the first of these, the sync state, where the swarmalators fall into sync dots spaced $\pi$ units apart  $x_i \in [C_1,C_1+\pi]$ and $\theta_i \in [C_2, C_2+\pi]$ \footnote{single cluster states are realized for some initial conditions}. Panel (b) plots the second, the phase wave, where the swarmalators positions and phases are linearly spaced and perfectly correlated $x_i = \pm \theta_i + const$. For the sake of completeness, panel (c) shows the model's third state\footnote{for the special cases $K'=0$ or $J'=0$ other states are realized, but those marginal cases don't interest us here} which is fully supported: asynchrony, where the swarmalators are scattered uniformly over both position and phase $x,\theta$.

The stability of asynchrony has been derived in the continuum limit, but the case of finite-$N$ remains elusive \cite{o2022collective}. For synchrony and the phases waves the situation is reversed: We have their stabilities for finite-$N$ \cite{o2022collective}, but the infinite-$N$ case is missing. This paper fills in this gap.

Why care about stability of the sync and phase waves states in the infinite-$N$ limit? We already know their stabilities for any finite $N$, after all, which can be made as large as we like. There are a few reasons. 

The first is theoretical interest. When crossing the bridge from `large-but-finite-$N$' to the continuum, one often encounters unexpected subtleties. Take the Kuramoto model. Two famous reductions take model's dimensionality from $N\gg1$ all the way down to 2 or 3 \footnote{a giant reduction for an otherwise unassailable many-body nonlinear dynamical system}. Yet one of these, the Watanabe-Strogatz transformation \cite{watanabe1993integrability}, works only for finite numbers of identical oscillators, while the other, the Ott-Antonsen ansatz \cite{ott2008low}, works only for an infinite number of nonidentical oscillators. The tension between these two results -- one working for finite $N$, the other for infinite $N$ -- created a slew of puzzles which ended up connecting coupled oscillator theory to diverse parts of mathematics such as group theory, harmonic analysis, and hyperbolic geometry \cite{lipton2021kuramoto,chen2017hyperbolic,engelbrecht2020ott}.

Could there be similar hidden connections in the discrete/continuum transition of the 1d swarmalator model? A firm understanding of the model's continuum behavior, which this paper provides, sets us up to explore this possibility.

The other more important reason to study the phase waves and sync states in the continuum limit is that these states pop up in other settings where finite-$N$ analyses tend to fail. Phase waves, for example, arise in the 1d model with disordered coupling \cite{o2022swarmalators}. Their stabilities are here unknown. Standard linearization for finite-$N$ doesn't work (characteristic equations become unwieldy) leaving a continuum analysis as the only viable option. This paper solves this open problem. 

\begin{figure}[hpt]
    \centering
    \includegraphics[width=1.0\columnwidth]{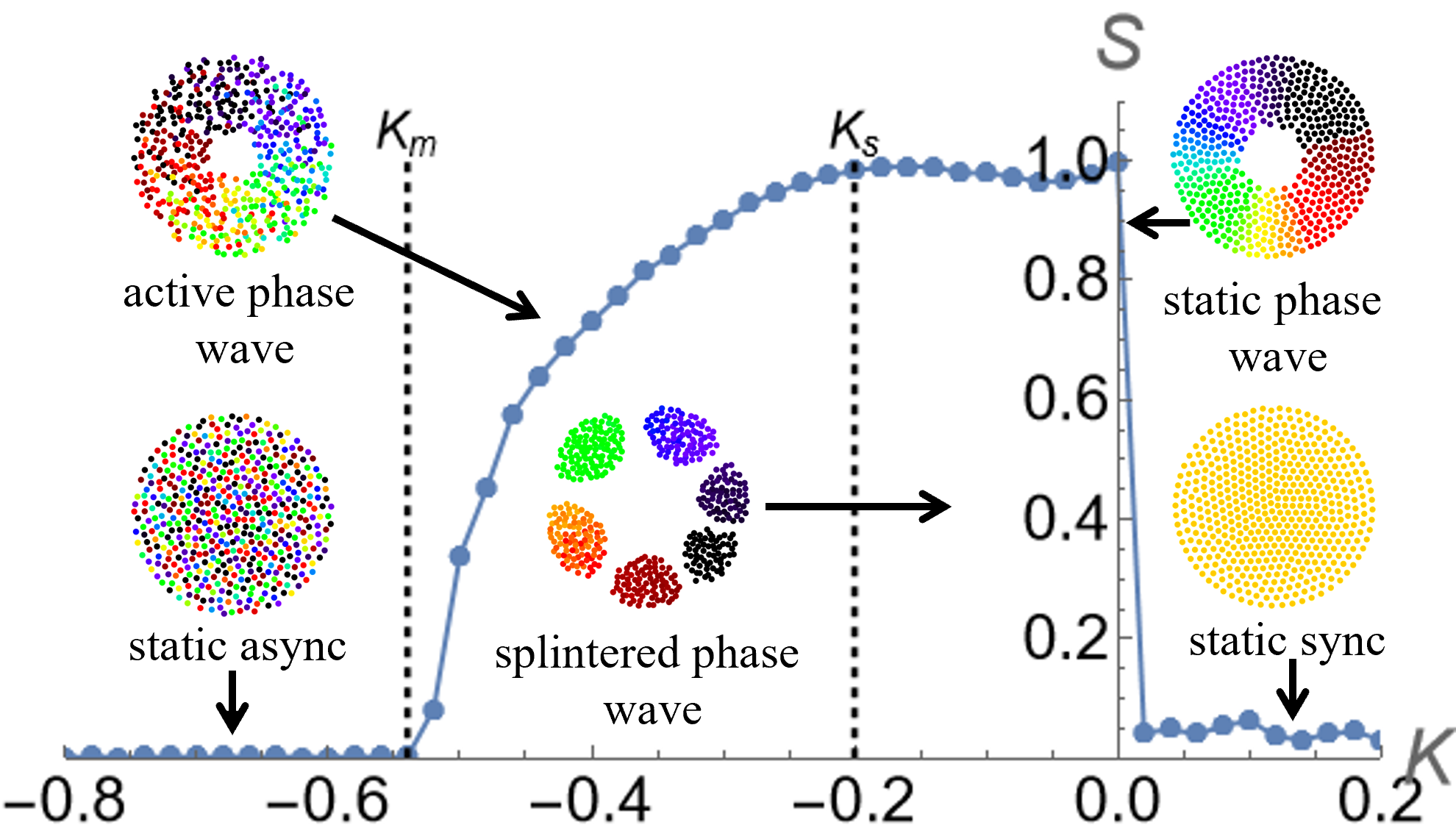}
    \caption{Rainbow order parameter $S(K)$ of the 2d swarmalator model \cite{o2017oscillators}. Insets show the model's collective states where swarmalators are represented as colored dots in the $(x,y)$ plane; the color denotes the swarmalators' phase. Static async: $(J,K) = (0.5,-0.8)$, active phase wave $(J,K) = (0.5,-0.4)$, splintered phase wave $(J,K) = (0.5,-0.1)$, static phase wave $(J,K) = (0.5,0)$, static sync $(J,K) = (0.5,0.2)$. In the three static states, swarmalators do not move in space or phase. In the splintered phase wave, each colored chunk is a vortex: the swarmalators oscillate in both space and phase. In the active phase wave, the oscillations are excited into rotations. The swarmalators split into two groups with counter-rotate in $x$ and $\theta$.} 
    \label{states-2d}
\end{figure}

And even more importantly, analogues of the phase wave and sync states also arise in the more realistic 2d swarmalator model \cite{o2017oscillators}. Figure~\ref{states-2d} shows this model's states along with their bifurcation structure by plotting its rainbow order parameter $S$ versus the coupling constant $K$ \footnote{The definition is $S := max(S_+, S-)$ with $S_{\pm} = \langle e^{i(\phi\pm\theta)}$ where $tan \phi = y/x$ is the spatial angle. For random initial conditions, the $\pm$ occur with equal probability so we use the max to break the degeneracy}. The disk-like static async and sync states are 2d analogues of the 1d model's sync dots. The static and active phase waves are analogues of the 1d phase wave.

A big open problem in the swarmalator field is to determine the stability of these states, as well as their critical parameter values -- to find the so-called melting point $K_m$ and splitting point $K_s$ (dashed lines in Figure~\ref{states-2d}) \cite{o2017oscillators, o2023solvable,gong2024approximating}. Finite-$N$ analyses are too difficult, because there are a giant number of fixed point solutions corresponding to a given disk or phase wave state. For infinite-$N$, the blocker is the states' compact support. For example, linearizing around the async state $\rho(\mathbf{x},\theta) = (2 \pi^2 R^2)^{-1} H(r-R)$, where $H(r)$ is Heaviside function's and $R$ is the disk's radius, produces eigenvalue equations of the form $\lambda b (\mathbf{x}) = H(r-R) \int b(\mathbf{y}) / |\mathbf{x}-\mathbf{y}| d \mathbf{y} + \delta(r-R) \int b(\mathbf{y})  (\mathbf{y}-\mathbf{x}) d \mathbf{y} $ \footnote{We are mixing cartesian coordinates $\mathbf{x}$ and polar coordinates $(r,\phi)$ for the spatial coordinates here ($\theta$ in the phase). We hope the meaning is clear.} These are hard to solve. Part of the difficulty is the Heaviside and delta functions that sit in front of the integrals (these are the mathematical signatures of the compact support). Such generalized functions takes us into a more advanced functional-analytic setting which is (to this author at least) non-standard.

The stability spectrum of the 1d phase wave has the same features as the above eigenvalue equation, but without the added complexity of the power law kernels. It is a 1d, simpler version of this 2d problem and so serves as a stepping stone to understanding 2d swarmalators more generally. 

That is what what makes the 1d model worth studying. It captures the essence of more complex swarmalator models, and, being so closely related to Kuramoto' model, might inherit some of that model's theoretical richness \cite{lipton2021kuramoto}.  


\section{Results} 

\subsection{Warm up problems}
We start with some warm up problems that expose the key features of our main problem. We share Mathematica notebooks with all our workings if you are curious about the details \cite{github}.

\subsubsection{Warm up 1}
Consider $N$ uncoupled particles moving on the real line according to
\begin{align}
    \dot{x}_i = - K x_i \label{warm1}
\end{align}
Here $i=1 \dots N$ and $x \in \mathbb{R}^1$. This has a fixed point at $x_i = 0$ for all $i$. The eigenvalues of the Jacobian at this point are
\begin{align}
    \lambda_{N} = -K \hspace{0.5 cm} \text{wm } N-1
\end{align}
where "wm" means \underline{w}ith \underline{m}ultiplicity $N$. The subscript $N$ remind us the $\lambda_{N}$ are for finite-$N$. Now, our goal is to repeat this analysis in the continuum limit, to find the eigenvalues $\lambda_{\infty}$.

Suppose $N$ is large enough so we can imagine the system as a fluid spread over the real line. Let $\rho(x,t)$ denote the density of particles. We interpret this in the Eulerian sense: $\rho(x,t) dx$ gives the fraction of particles between position $x$ and $x+dx$ at time $t$. This $\rho$ then satisfies the continuity equation:
\begin{align}
    \dot{\rho} + \frac{\partial}{\partial x} (v \rho) = 0
\end{align}
The velocity is given by the ODE above
\begin{align}
    v(x) = - K x
\end{align}
The interpretation here is that a test particle at position $x$ has velocity $-Kx$. Consider perturbations around the `fixed point' in density space $\rho_0(x) = \delta(x)$
\begin{align}
\rho(x) = \rho_0(x) + \epsilon \rho_1(x) \\
\rho(x) = \delta(x) + \epsilon \rho_1(x)
\end{align}
Normalization requires 
\begin{align}
& \int \rho(x) dx = 1 \\
& \int \rho_0 + \epsilon \rho_1(x) = 0 \\
& 1 + \epsilon \rho_1(x) = 0 \\
& \int \rho_1(x) = 0    
\end{align}
Now we plug this into the continuity equation and collect terms at $O(\epsilon)$. The result is
\begin{align}
    \rho_1'(x) = K \Big( \rho_1(x) + x \rho_1'(x) \Big)
\end{align}
Letting $\rho_1(x) = \lambda b(x)$ we get
\begin{align}
    & \lambda b(x) = K \Big( b(x) + x b'(x) \Big) \label{ee_w1} \\
    & \int b(x) dx = 0
\end{align}
where the second equation requotes the normalization condition. This is the eigenvalue equation we must solve. It has form $\lambda b = L b$, where $L$ is a linear operator. This operator has both a continuous and a discrete spectrum.

\textit{Discrete spectrum}. To find the discrete spectrum, we seek all \(\lambda\) such that the eigenvalue equation \eqref{ee_w1} has normalizable solutions \(b(x)\) in \(L^2\). One natural guess is \(b(x) = x^n\), which satisfies the equation for specific \(n\). However, these solutions do not belong to \(L^2\) because they diverge at \(x = \pm \infty\), rendering them unphysical.

Given that \(\rho_0(x)\) is a delta function, another natural guess is delta functions and their derivatives (which are in $L^2$). Consider the $n$-th derivative
\begin{align}
b(x) = \delta^{(n)}(x)    
\end{align}
where normalization requires $n>1$.  Plugging this into Eq.~\eqref{ee_w1} gives
\begin{align}
& \lambda \delta^{(n)}(x) = K( \delta^{(n)}(x)  - x \delta^{(n+1)}(x)) \\
& \lambda \delta^{(n)}(x) = - n K\delta^{(n)}(x) 
\end{align}
where we used the identity $x \delta^{(n+1)}(x) = -(n+1) \delta^{(n)}(x)$. We see the spectrum is
\begin{align}
\lambda_{\infty} = -n K \hspace{0.5 cm} n>0, \; \; n \in \mathbb{Z}^+
\end{align}
where the $\infty$ subscript reminds us these are the eigenvalues of the continuum limit.

This is our desired solution. It says the $\rho_0(x) = \delta(x)$ is stable for all $K>0$. This agrees with the finite-$N$ analysis, but notice the magnitudes of $\lambda_{\infty}$ and $\lambda_N$ are different: the former has a scaling factor of $n$. This feature of the $(\lambda_N, \lambda_{\infty})$ pair, different scaling factors but identical stability regions, recurs throughout our analysis. 

\textit{Continuous spectrum}. This is defined as the set of all $\lambda$ for which the operator $L - \lambda I$ is \textit{not} surjective. So we consider
\begin{align}
(1-\lambda)b(x) + x b'(x) = f(x)
\end{align}
For some $f(x)$. Surjectivity means we can always find a $b^*(x)$ that produces any desired $f^*(x)$. If there is a $\lambda$ for which this is \textit{not} possible, then that $\lambda$ is in the continuous spectrum. You can see that there are no such $\lambda$ in the equation above. 

You might be tempted to think $\lambda=1$ is a contender. But in that case we get
\begin{align}
x b'(x) &= f(x) \\
b(x) &= \int f(x') / x' dx
'\end{align}
So we see we can find a $b(x)$ that satisfies a desired $f(x)$.

To cut the the chase, for all the stability problems we study in this paper, there will be no continuous spectra. For pedagogical purposes, we give an example of an operator \textit{with} a continuous spectrum in the \textit{Appendix}.

\subsubsection{Warm up 2}
Now consider $N$ \textit{coupled} particles moving on the real line
\begin{align}
    \dot{x}_i = \frac{K}{N} \sum_j (x_j-x_i) \label{warm2}
\end{align}
The finite-$N$ eigenvalues are
\begin{align}
    & \lambda_{N,0} = 0, \hspace{0.5 cm} \text{wm } 1 \label{l1a}  \\
    & \lambda_{N,1} = -K, \hspace{0.5 cm} \text{wm } N-1 \label{l2a}
\end{align}
The zero eigenvalues comes from the translational symmetry. Moving $x_i \rightarrow x_i + const$ doesn't change the governing ODEs since the constant cancels out in the difference $x_j-x_i$. To find $\lambda_{\infty}$, we follow the same procedure as before. The only difference is the velocity has both an $O(1)$ and $O(\epsilon)$ term
\begin{align}
v(x) &= \int (x'-x) \rho(x') dx' \\
v(x) &= \int (x'-x) ( \rho_0(x') + \epsilon \rho_1(x',t) ) dx' \\
v(x) &= \int (x'-x) ( \delta(x') + \epsilon \rho_1(x',t) ) dx' \\
v(x) &= -x + \epsilon(\mu_1 - x \mu_0 ) \\
v(x) &= -x + \epsilon \mu_1
\end{align}
where $\mu_n := \int x'^n \rho_1(x') dx'$ is the $n$-th moment of the perturbed density. Recall normalization requires $\mu_0 = 0$ which we used in the second to last line. Plugging into the continuity equation gives
\begin{align}
    & \lambda b(x) = K \Big( b(x) + x b'(x) + \mu_1 \delta'(x) \Big) \label{ee_w2} \\
    & \int b(x) dx = 0 \\ 
    & \mu_1 =  \int x b(x)  dx 
\end{align}
The effect of the new $\mu_1$ term is to shift the starting point of the spectrum to $n=2$. To see this, we insert a \textit{delta series} ansatz for $b(x)$
\begin{align}
    b(x) = \sum_{n=1} c_n \delta^{(n)}(x)
\end{align}
into the eigenvalues equation Eq.~\eqref{ee_w2} and then match coefficients of the delta functions. This gives a set of simultaneous equations
\begin{align}
    c_1 &= -2 k c_1 \\
    c_{n>1} &= -n K c_{n>1}
\end{align}
(In warm up 1, the $c_1$ equation was \textit{not} distinguished; it has coefficient $-1$ instead of $-2$ on the rhs). The solution is
\begin{align}
\lambda_{\infty} = -n K \hspace{0.5 cm} n>1, \; \; n \in \mathbb{Z}^+
\end{align}
where we remind you the series starts at $n=2$. As before, $(\lambda_N, \lambda_{\infty})$ have different scaling factors but describe the same stability region $K>0$. 

For the same reasons as last time, the continuous spectrum does not exist.

\subsubsection{Warm up 3}
Finally, consider the homogeneous Kuramoto model
\begin{align}
    \dot{\theta}_i &= \omega + \frac{K}{N} \sum_j \sin (\theta_j - \theta_i) \\
    \dot{\theta}_i &= \omega +  K r  \sin (\psi - \theta_i)
\end{align}
where $r e^{i \psi} = \langle e^{i \theta} \rangle$ is the order parameter. We can set $\omega = 0$ without loss of generality. This implies the average phase is constant $\dot{\langle \theta \rangle} = 0$, which we get from summing over all $i$ on both sides of the equation. We can assume the initial $\langle \theta \rangle$ is zero also without loss of generality which means $\psi = 0$. Putting this altogether gives an especially simple equation
\begin{align}
    \dot{\theta}_i = -K r  \sin \theta_i \label{vel}
\end{align}
We are interested in the stability of the sync state $\theta_i = C = 0$ (the $0$ coming from the assumptions we made above). 

The finite-$N$ eigenvalues are
\begin{align}
    & \lambda_0 = 0, \hspace{0.5 cm}  \text{wm } 1 \label{l1aa}  \\
    & \lambda_1 = -K, \hspace{0.5 cm}  \text{wm } N-1 \label{l2aa}
\end{align}
like before. The continuum analysis induces a velocity
\begin{align}
    v = -L  \sin \theta + \epsilon k r_1(t) \sin( \psi_1(t) - \theta)  \label{x2}
\end{align}
where  
\begin{align}
    r_1 e^{\psi_1} = \int e^{i \theta'} \rho_1(\theta', t) d \theta'
\end{align}
are the perturbed order parameters (in the sense it;s the first mode of the perturbed density $\rho_1$). Notice that the $v_0$ term is non-zero; this contrasts with the stability analysis of the async state $\rho_0 = 4\pi^{-2}$, a much more common and well understood calculation \cite{strogatz1991stability}, in which $v_0 = 0$.

The eigenvalue equation becomes
\begin{align}
 \lambda b = K \Big( b \cos \theta +  \sin \theta b_{\theta} -  r_1 \sin \psi_1 \delta' (\theta) \Big) \label{ee1}
\end{align}
We again use a delta series
\begin{align}
    b(\theta) = \sum_{n=1} c_n \delta^{(n)}(\theta) \label{x4}
\end{align}

Surprisingly, the spectrum is the same as Warm up 2
\begin{align}
\lambda &= 0 \\
\lambda &= - n K,  \hspace{0.5 cm} n>1, \; \; n \in \mathbb{Z}^+
\end{align}
As before, the continuous spectrum does not exist.

\subsection{Simplifying the 1d swarmalator model}
With the warm up problems under our belt, we are now ready to tackle the 1d swarmalator model. First we make some simplifications. We are interested in the case of identical swarmalators which all have the same natural frequency which can be set to zero via a change of frame $\omega_i = \nu_i = 0$. Then we move to sum/difference coordinates $(\xi, \eta) := (x+\theta, x-\theta)$ which makes the model simpler
\begin{align}
\dot{\xi}_i &=  K r \sin(\phi - \xi) + J s \sin(\psi - \eta)  \\
\dot{\eta}_i &= J r \sin(\phi - \xi) + K s \sin(\psi - \eta)  
\end{align}
where
\begin{align}
& r e^{i \phi}= \langle e^{i \xi} \rangle \\
& s e^{i \psi}  = \langle e^{i \eta} \rangle  \\
& (J,K) = ( (J'+K')/2, (J'-K')/2)
\end{align}
Like before, the average coordinates are constant throughout the dynamics and thus can be set to zero without loss of generality $\langle \xi \rangle = \langle \eta \rangle = 0$. This means the phases of the order parameters are also zero $\psi = \phi =0$
\begin{align}
\dot{\xi}_i &= - K r \sin \xi - J s \sin \eta  \\
\dot{\eta}_i &= -J r \sin \xi - K s \sin \eta  
\end{align}
The model is now especially simple. It is a pair of linearly coupled Kuramoto models, with the rainbow order parameters $r,s$ generalizing the Kuramoto order parameter $ \langle e^{i\theta} \rangle$ \cite{kuramoto2003chemical}. In fact, in the special case $J=0$, the model decouples into two independent Kuramoto models in the coordinate $\xi$. The case $K=0$ is also interesting -- the model becomes Hamiltonian -- but that does not concern us here.

One benefit of the $(\xi, \eta)$ frame is phase waves are especially simple to describe in it. There is a $\eta$-wave, where the $\eta$ are perfectly in sync and the $\xi$'s are distributed uniformly: $\rho(\xi,\eta) = \delta(\eta) / 2\pi$ and an $\xi$-wave, where the reverse happens. Both kinds of phase wave are equally likely due to the model's $(\xi,\eta) \rightarrow (\eta, \xi)$ symmetry. We analyze the $\eta$-wave without loss of generality.

Our goal is thus to find the stability of the phase wave in the infinite-$N$ limit by linearizing around the density
\begin{align}
\rho(\xi,\eta) = \frac{\delta(\eta)}{2\pi}    
\end{align}

\subsection{Stability of phase wave in continuum limit}
The finite-$N$ eigenvalues have been worked out previously \cite{o2022collective} and are
\begin{align}
    & \lambda_{N,0} = 0, \hspace{0.5 cm} \text{wm }  N-1 \label{l1}  \\
    & \lambda_{N,1} = -K, \hspace{0.5 cm} \text{wm } N-3 \label{l2} \\
    & \lambda_{N,2} = \frac{1}{4} \Big( -K \pm \sqrt{ 9 K^2 - 8 J^2} \Big), \hspace{0.5 cm}  \text{wm } 2 \label{l3}
\end{align}
Now consider small perturbations $\rho_1$ around the $\eta$-wave
\begin{align}
\rho(\xi,\eta) &= \rho_0(\xi,\eta) + \epsilon \rho_1(\xi,\eta) \\
\rho(\xi,\eta) &= \frac{\delta(\eta)}{2\pi} + \epsilon \rho_1(\xi,\eta) \label{rho}
\end{align}
The velocity fields $v$ become
\begin{align}
& v_{\xi} = - J \sin \eta + \epsilon(K r_1 \sin(\phi_1 - \xi) + J s_1 \sin(\psi_1 - \eta)) \label{v1} \\
& v_{\eta} = - K \sin \eta + \epsilon(J r_1 \sin(\phi_1 - \xi) + K s_1 \sin(\psi_1 - \eta)) \label{v2}
\end{align}
where
\begin{align}
r_1 e^{i \psi_1} = \int e^{i \xi} \rho_1(\xi,\eta) d \xi d \eta \label{r1} \\
s_1 e^{i \phi_1} = \int e^{i \eta} \rho_1(\xi,\eta) d \xi d \eta \label{s1}
\end{align}
The discrete spectrum is
\begin{align}
   \lambda b &= K b \cos \eta + \sin \eta \left( K b_{\eta}  + J b_{\xi} \right) - \frac{s_1}{2 \pi} K \sin \phi_1 \delta'(\eta) \nonumber \\
   &+\frac{r_1}{2 \pi} \Big( K \delta(\eta) \cos(\xi - \phi_1) + J \sin(\xi - \phi_1) \delta'(\eta) \Big) \label{ee}
\end{align}
This time, the right ansatz is a Fourier series over $\xi$
\begin{align}
    b(\xi,\eta) = \frac{1}{2\pi} \Big( a_0(\eta) + \cos(\xi) a_1(\eta) + \sin(\xi) b_1(\eta) \Big)
\end{align}
(higher order terms are unnecessary, since the rhs of Eq.~\eqref{v1}, ~\eqref{v2} contain only first and zeroth order harmonics in $\xi$). The Fourier coefficients are delta series
\begin{align}
    & a_0(\eta) = \sum_{n=1} c_{n} \delta^{(n)}(\eta) \\
    & a_1(\eta) = \sum_{n=0} d_{n} \delta^{(n)}(\eta) \\
    & b_1(\eta) = \sum_{n=0} e_{n} \delta^{(n)}(\eta)
\end{align}
Importantly, normalization only requires the zeroth derivative of the $a_0(\eta)$ term to be zero: $\int b(\xi, \eta) d \xi d \eta = 0 \Rightarrow \int a_0(\eta) d \eta = 0$. So the summands in $a_1(\eta), b_1(\eta)$ start at $n=0$ and thus include $\delta(\eta)$. Inserting the ansatz into the eigenvalue equation and simpliftying yields a complex expression $E$. We project this onto the relevant Fourier modes separately.

\textit{Zeroth mode}. Integrating $\int E d \xi$ yields an set of simultaneous equations \cite{github} whose solution are
\begin{align}
    & \lambda_{\infty,0} = 0 \\
    & \lambda_{\infty,1} = -n K, \hspace{0.5 cm} n > 2 \; \; n \in \mathbb{Z}^+
\end{align}
which echo our previous findings.

\textit{First modes}. Integrating $ \int E \cos \xi d \xi, \int E \sin \xi d \xi $ yields equations of form
\begin{align}
 \frac{d_0 \lambda }{2} &= \frac{d_0 K}{4}-\frac{e_1 J}{2}+\frac{e_3 J}{2} \\
 \frac{d_1 \lambda }{2} &= -\frac{d_1 K}{2}+\frac{d_3 K}{2}-\frac{e_0 J}{4}-e_2 J + \dots \\
 \frac{d_2 \lambda }{2} &= d_2 (-K)-\frac{3 e_3 J}{2} \\
 \frac{d_3 \lambda }{2} &= -\frac{3 d_3 K}{2} + \dots \\
 \frac{e_0 \lambda }{2} &= \frac{d_1 J}{2}-\frac{d_3 J}{2}+\frac{e_0 K}{4} \\
 \frac{e_1 \lambda }{2} &= \frac{d_0 J}{4}+d_2 J-\frac{e_1 K}{2}+\frac{e_3 K}{2} + \dots \\
 \frac{e_2 \lambda }{2} &= \frac{3 d_3 J}{2}-e_2 K + \dots \\
 \frac{e_3 \lambda }{2} &= -\frac{3 e_3 K}{2} \\
\end{align}
where the $\dots$ denote terms with $d_{i>3}, e_{i>3}$. The solutions for $c_i, d_i, e_i$ are rather long and unimportant for stability concerns, so we don't display them (see \cite{github}). The solutions for $\lambda$ are however clean
\begin{align}
    & \lambda_{\infty, 0} = 0 \\
    & \lambda_{\infty, 1} = -n K, \hspace{0.5 cm} n > 2, \; \; n \in \mathbb{Z}^+ \\
    & \lambda_{\infty, 2} =  \frac{1}{4} \Big( -K \pm \sqrt{9 K^2 - 8 J^2} \Big)
\end{align}
These are our final answers. The $\lambda_{\infty,0}, \lambda_{\infty,2}$ match $\lambda_{N,0}, \lambda_{N,2}$ exactly (Eq.~\eqref{l1},~\eqref{l2}), but the $\lambda_{\infty,1} = -nK$ again has the scaling factor $n$.

Similar considerations as for the warm up cases show there is no continuous spectrum. 

This completes our analysis of the phase wave.

\subsection{Stability of phase wave for $K_j$ coupling}
Phase waves occur in a generalized 1d swarmalator model where the uniform couplings have been replaced with random quantities like so
\begin{align}
    \dot{x}_i &= \frac{1}{N} \sum_j J_j' \sin(x_j - x_i) \cos(\theta_j - \theta_i) \\
    \dot{\theta}_i &= \frac{1}{N} \sum_j  K_j' \sin(\theta_j - \theta_i) \cos(x_j - x_i) 
\end{align}
The effect is to make the swarmalator interactions non-reciprocal, which arises commonly in biology. The stability of the phase wave is unknown; finite-$N$ analysis were unsuccessful. Our continuum analysis can however handle this case. 

The main difference is the density now depends on the random couplings, $\rho(\xi,\eta,J,K,t)$. The interpretation is that $\rho(\xi,\eta,J,K,t) d \xi d\eta dK dJ$ gives the fraction of swarmalators with phases between $(\xi,\eta)$ and $(\xi+d\xi,\eta+d\eta)$ and couplings between $(J,K)$ and $(J+dJ, K+dK)$ at time $t$. The velocities become
\begin{align}
& v_{\xi} = - r_{K} \sin \xi - s_{J} \sin \eta \\
& v_{\eta} = - r_{J} \sin \xi - s_{K} \sin \eta 
\end{align}
Notice the new order parameter are now coupling-weighted
\begin{align}
    r_{(K,J)} &=  \int (J,K) e^{i \xi} \rho(\xi,\eta,J,K) g(J) h(K) d\xi d\eta dJ dK \\
    s_{(K,J)} &= \int (J,K) e^{i \eta} \rho(\xi,\eta,J,K) g(J) h(K) d\xi d\eta dJ dK
\end{align}
where we use $(K,J)$ notation to denote either $J$ or $K$. The $g(J), h(K)$ are the densities of $J,K$. Plugging in the $\rho = \rho_0 + \epsilon \rho_1$ ansatz yields
\begin{align}
& v_{\xi} = - \mu_J \sin \eta + \epsilon \Big( r_{K,1} \sin(\phi_{K,1} - \xi) + s_{J,1} \sin(\psi_{J,1} - \eta) \Big) \label{v1bb} \\
& v_{\eta} = - \mu_K \sin \eta + \epsilon \Big( r_{J,1} \sin(\phi_{J,1} - \xi) + s_{K,1} \sin(\psi_{K,1} - \eta) \Big) \label{v2bb}
\end{align}
We see the difference is the appearance of $\textit{mean}$ couplings $\mu_J, \mu_K$ in the unperturbed piece of the velocity. The discrete spectrum becomes
\begin{align}
   \lambda b = & \mu_K b \cos \eta + \sin \eta \left( \mu_K b_{\eta}  + \mu_J b_{\xi} \right) - \frac{s_{K,1}}{2 \pi} \sin \phi_{K,1} \delta'(\eta) \nonumber \\
   &+\frac{1}{2 \pi} \Big( r_{K,1} \delta(\eta) \cos(\xi - \phi_{K,1}) + r_{J,1} \sin(\xi - \phi_{1,J}) \delta'(\eta) \Big) \label{ee3}
\end{align}
where, recall, $b=b(\xi,\eta,K,J)$. The ansatz is the same as before
\begin{align}
b(\xi,\eta,J,K) = \frac{1}{2\pi} \Big( a_0(\eta,K,J) + a_1(\eta,K,J) 
\cos \xi \\ + b_1(\eta,K,J) \sin \xi  \Big)
\end{align}

\textit{Zeroth mode}. Projecting onto the zeroth mode yields
\begin{align}
   \lambda a_0(\eta,K,J) = & \mu_K a_0 \cos \eta + \mu_K a_{0,\eta} \sin \eta - \frac{s_{K,1}}{2 \pi} \sin \phi_{K,1} \delta'(\eta)
   \label{eec}
\end{align}
We use the same ansatz for $a_0(\eta,J,K) = \sum_{n=1} c_n \delta^{(n)}(\eta)$. Importantly, the $c_n$ do \textit{not} depend on the $(J,K)$. The expression for the order parameter becomes
\begin{align}
    s_{k,1} &= \int K e^{i \eta} b(\xi,\eta,J,K) g(J) h(K) d \xi d \eta dJ dK \\
      &= \int K e^{i \eta} a_0(\eta,J,K) g(J) h(K) d \eta dJ dK \\
      &= \int K (-i c_1 + c_2 + i c_3 + c_4 + \dots) g(J) h(K)  dJ dK \\
      &= \mu_K (-i c_1 + c_2 + i c_3 + c_4 + \dots) 
\end{align}
Plugging this expression for $s_{K,1}$ along with the ansatz for $a_0(\eta)$ into the Eq.~\eqref{eec} and matching coefficients of the delta functions leads to a set of simultaneous equations like before. The solution is
\begin{align}
    & \lambda_{\infty, 0} = 0 \\
    & \lambda_{\infty, 1} = -n \mu_K, \hspace{0.5 cm} n > 2, \; \; n \in \mathbb{Z}^+
\end{align}

\textit{First modes}. The procedure is the same here so we just quote the result
\begin{align}
    & \lambda_{\infty,0} = 0 \\
    & \lambda_{\infty, 1} = -\frac{n}{2} \mu_K, \hspace{0.5 cm} n > 2 , \; \; n \in \mathbb{Z}^+ \\
    & \lambda_{\infty, 2} =  \frac{1}{4} \Big( -\mu_{K} \pm \sqrt{9 \mu_{K}^2 - 8 \mu_{J}^2} \Big)
\end{align}
Aside from the new divisor of $2$ in $\lambda_{\infty,1}$, the only difference between the $K_j$ and $K$ coupling is that the $(K,J)$ get replaced by their mean values. This confirms the conjecture made in \cite{o2022swarmalators}. 

Notice that these stability results hold for \textit{any} distributions $g(J), h(K)$. This is quite surprising. The only features of these densities that influence the dynamics are their mean values.

\subsection{Stability of sync states in continuum limit}
Finally, we close with an analysis of the sync state defined by $\xi_i = C_1$ and $\eta_i = C_2$ for all $i$. We set $C_1,C_2 = 0$ without loss of generality. Then the density we perturb around is
\begin{align}
    \rho(\xi,\eta) = \delta(\xi) \delta(\eta)
\end{align}
The density and velocity fields decompose as
\begin{align}
& \rho(\xi,\eta) = \delta(\xi) \delta(\eta) + \epsilon \eta(\xi,\eta,t) \\
& v_{\xi} = -K \sin \xi - J \sin \eta + \epsilon(K r_1 \sin(\phi_1 - \xi) + J s_1 \sin(\psi_1 - \eta)) \\
& v_{\eta} = -J \sin \xi - K \sin \eta + \epsilon(J r_1 \sin(\phi_1 - \xi) + K s_1 \sin(\psi_1 - \eta)) 
\end{align}
Subbing this into the continuity equation and expanding gives
\begin{align}
\lambda b(\xi,\eta) = & b_{\eta}(\xi,\eta) (J \sin \xi + K \sin \eta ) \nonumber \\
& + b_{\xi}(\xi,\eta) (K \sin \xi + J \sin \eta) \nonumber \\
& + K (\cos \xi + \cos \eta ) b(\xi,\eta) \nonumber \\
& + r (K \delta(\eta) \delta'(X) \sin(\xi - \phi_1) \nonumber \\
& + J \delta(\xi) \sin(\xi - \phi_1) \delta'(\eta) \nonumber \\
& + K \delta(\xi) \delta(\eta) \cos(\phi_1)) \nonumber \\
& + s (K \delta(\xi) \delta'(Y) \sin(\eta - \psi_1) \nonumber \\
& + J \delta(\eta) \delta'(X) \sin(\eta - \psi_1) \nonumber \\
& + K \delta(\xi) \delta(\eta) \cos(\psi_1)).
\end{align}
The ansatz 
\begin{align}
 b(\xi,\eta) = \sum_{\substack{i=0, j=0 \\ (i,j) \neq (0,0)}} \delta^{i}(\xi) \delta^{(j)}(\eta)   
\end{align}
eventually yields
\begin{align}
& \lambda_{\infty,0} = 0 \\
& \lambda_{\infty,1} = -n (J+K) \hspace{0.5 cm} n>2, \; \; n \in \mathbb{Z}^+ \\
& \lambda_{\infty,2} = -m (J-K) \hspace{0.5 cm} m>2, \; \; m \in \mathbb{Z}^+ 
\end{align}
which results in the same stability region as the finite-$N$ case \cite{o2022collective}.

\section{Discussion}
\begin{table}[t!]
\centering
\rowcolors{1}{gray!25}{white}
\begin{tabular}{|l|c|c|}
\hline
\rowcolor{gray!50}
\textbf{State (identical frequencies)} & \textbf{Stability finite-$N$} & \textbf{Stability Infinite-$N$} \\ \hline
Async (1d) & \cellcolor{yellow!50}?? & \cellcolor{green!30}\checkmark \\ \hline
Sync (1d) & \cellcolor{green!30}\checkmark & \cellcolor{green!30}\checkmark \\ \hline
Phase wave (1d) & \cellcolor{green!30}\checkmark & \cellcolor{green!30}\checkmark \\ \hline
\rowcolor{white} 
& & \\ \hline
Static async (2d) & \cellcolor{yellow!50}?? & \cellcolor{yellow!50}?? \\ \hline
Static sync (2d) & \cellcolor{yellow!50}?? & \cellcolor{yellow!50}?? \\ \hline
Static phase wave (2d) & \cellcolor{yellow!50}?? & \cellcolor{yellow!50}?? \\ \hline
Active phase wave (2d) & \cellcolor{yellow!50}?? & \cellcolor{yellow!50}?? \\ \hline
Splintered phase wave (2d) & \cellcolor{yellow!50}?? & \cellcolor{yellow!50}?? \\ \hline
\end{tabular}
\caption{What we know about the stability of swarmalator models (with identical natural frequencies and uniform couplings). For the 1d model, what's missing are results for finite populations of swarmalators in the async state. For the 2d model, the field is wide open: the stabilities of \textit{all} of its states are a mystery.} \label{table}
\end{table}
We have found the stability spectra for the sync and phase wave states of the 1d swarmalator model in the continuum limit. From a stability perspective, our understanding of the model is almost complete. Table~\ref{table} shows all that's left is the stabilities of the async state for finite-$N$. We are working on these and think we have it figured out. There are some surprises -- we've seen hints of glassiness for small $N$ and think we have a way to characterize it analytically -- which we will soon share.

The significance of all this -- that is, of a well understood 1d model -- is it sets us up to tackle the harder 2d swarmalator model whose stabilities and bifurcations are a complete mystery. Those are the really juicy open problems. In the eight years since they were reported \cite{o2017oscillators}, no one has managed to touch them. With our new understanding of linear operators containing generalized functions, however, we think we are close. We are still working out the details -- for instance, one needs to assign values to \textit{products} of generalized functions like $\delta''(r-R) H(R-r)$ which requires advanced distribution theory \cite{aragona1991intrinsic} -- but we plan to share what we have with the community soon. 

As for future work, recall the bulk of swarmalator studies focus on the simple case of units with \textit{identical} natural frequencies \cite{blum2024swarmalators,smith2024swarmalators,o2018ring,lizarraga2020synchronization,lizarraga2024order,ha2019emergent,hong2021coupling,adorjani2024motility,senthamizhan2024data,degond2023topological}. The more realistic case of dissimilar frequencies are an open story. Although we do know some things about the 1d model in this case. We know it has analogues of the async, sync, and phase wave states, and have managed to pull out their existence criteria \cite{yoon2022sync}. We know it has strange mixed state, halfway between the phase wave and sync state, where the swarmalators hop erratically between two synchronous clumps. But other than that, the dynamics of swarmalators with nonidentical frequencies are a wide open field, ripe for discovery and exploration.

\section{Code availability}
Github repository with all code available here \cite{github}.

\section{Ackowledgement}
We thank Rennie Mirollo for suggesting to us the warm up problems.


\section{Appendix}
We walk through a simple example of a linear operator \textit{with} a continuous spectrum (recall that all the examples in the main text did not have one). Our goal is to give some beginning intuition for how they work. For a proper introduction, see Beck's excellent notes on the linear stability of PDEs. \cite{beck}

Consider the linear operator
\begin{align}
    L_1 b(\omega) &= i \omega b(\omega) + \int_{-\infty}^{\infty} \frac{b(\nu)}{\nu^2 + \lambda^2} g(\nu) dy
\end{align}
where $g(\nu)$ is a probability density. This operator arises when analyzing the stability of the incoherent state in the Kuramoto model \cite{strogatz1991stability}. The definition of the continuous spectrum is the set of $\lambda$ for which the operator $L_1-\lambda I$ is \textit{not} surjective. The standard move is to consider the equation
\begin{align}
    (i \omega -\lambda) b(\omega) + \int_{-\infty}^{\infty} \frac{b(\nu)}{\nu^2 + \lambda^2} g(\nu) dy = f(\omega)
\end{align}
Surjectivity means that for every \( f(\omega) \) on the right-hand side, there exists at least one function \( b(\omega) \) on the left-hand side that satisfies the equation. The question is: for what values of \( \lambda \) does this property break down? Consider $\lambda = i \omega$. The leads to the equation
\begin{align}
\int_{-\infty}^{\infty} \frac{b(\nu)}{\nu^2 + \lambda^2} g(\nu) dy = f(\omega)
\end{align}
Here, no matter what \( b(\omega) \) we insert into the (definite) integral, a constant comes out. This implies that the operator \( L_1 - \lambda I \) no longer distinguishes between different inputs \( b(\omega) \), meaning it is no longer surjective. Therefore, \( \lambda = i \omega \) belongs to the continuous spectrum, as it fails to produce unique outputs for every input function \( b(\omega) \).

\end{document}